  \documentclass[structabstract]{aa}
\usepackage{graphicx}
\usepackage{txfonts}
\usepackage{natbib}
\usepackage{longtable}

\bibpunct[]{(}{)}{;}{a}{}{,}
\begin{document}
   \title{Rotational spectra of isotopic species of methyl cyanide, CH$_3$CN, in 
          their $\varv _8 = 1$ excited vibrational states}

   \author{Holger~S.~P. M{\"u}ller\inst{1}
           \and
           Brian~J. Drouin\inst{2}
           \and
           John~C. Pearson\inst{2}
           \and
           Matthias~H. Ordu\inst{1}
           \and
           Nadine Wehres\inst{1}
           \and
           Frank Lewen\inst{1}
          }

   \institute{I.~Physikalisches Institut, Universit{\"a}t zu K{\"o}ln,
              Z{\"u}lpicher Str. 77, 50937 K{\"o}ln, Germany\\
              \email{hspm@ph1.uni-koeln.de}
         \and
              Jet Propulsion Laboratory, California Institute of Technology, 
              Mail Stop 183-301, 4800 Oak Grove Drive, 
              Pasadena, CA 91011-8099, USA}

   \date{Received 20 October 2015 / Accepted 27 November 2015}

  \abstract
{Methyl cyanide is an important trace molecule in space, especially in 
star-forming regions where it is one of the more common molecules used 
to derive kinetic temperatures.}
{We want to obtain accurate spectroscopic parameters of minor isotopologs 
of methyl cyanide in their lowest excited $\varv _8 = 1$ vibrational states 
to support astronomical observations, in particular, with interferometers 
such as ALMA.}
{The laboratory rotational spectrum of methyl cyanide in natural isotopic 
composition has been recorded from the millimeter to the terahertz regions.}
{Transitions with good signal-to-noise ratios could be identified for the 
three isotopic species CH$_3^{13}$CN, $^{13}$CH$_3$CN, and CH$_3$C$^{15}$N 
up to about 1.2~THz ($J'' \le 66$). Accurate spectroscopic parameters were 
obtained for all three species.}
{The present data were already instrumental in identifying $\varv _8 = 1$ lines 
of methyl cyanide with one $^{13}$C in IRAM 30~m and ALMA data toward Sagittarius 
B2(N).}

\authorrunning{H.~S.~P. M{\"u}ller}
\titlerunning{rotational spectra of vibrationally excited isotopic CH$_3$CN}

\maketitle
\hyphenation{cyano-methane}
\hyphenation{Na-ti-o-nal}
\hyphenation{con-tri-bu-tion}
\hyphenation{Sym-po-si-um}
\hyphenation{Co-lum-bus}

%

\section{Introduction}
\label{intro}

The degree of molecular complexity in the interstellar medium (ISM) is particularly large 
in the warm and dense parts of star-forming regions known as ``hot cores'' or ``hot corinos''. 
A large number of organic molecules with up to 12 atoms have been detected thus far, for 
example, in the high-mass star-forming region Sagittarius (Sgr for short) B2(N) 
\citep{det-PrCN_EtFo,SgrB2-survey_2013,i-PrCN_det_2014}. The Orion~KL region is another 
high-mass star-forming region in which molecular complexity has been studied extensively 
(e.g., \citealt{search_EME_n-PrOH_2015}, and references therein). IRAS 16293-2422 
(e.g., \citealt{IRAS16293_glycald_2012,IRAS16293-ALMA-SV_2012}, and references therein) 
and NGC 1333-IRAS2A \citep{NGC1333-IRAS2A_survey_2014} are corresponding examples 
of low-mass star-forming regions. The Molecules in Space 
webpage\footnote{https://cdms.astro.uni-koeln.de/classic/molecules} of the Cologne Database for 
Molecular Spectroscopy, CDMS \citep{CDMS_1,CDMS_2} lists molecules detected in various 
astronomical sources with detailed information in most cases.

Molecular complexity in the ISM is not only manifest in increasingly larger molecules, 
but also in excited vibrational states or minor isotopologs of abundant smaller molecules. 
Corresponding laboratory rotational data are not only necessary to be able to avoid confusing 
such lines with those of larger and rarer molecules, but these data also have certain 
diagnostic values. Transitions in excited vibrational states may be used to derive a physical 
model of a particular source, as done, for example, in the case of CRL~618 and vibrationally 
excited HC$_3$N isotopologs \citep{CRL618-model_2003}. Ratios of isotopic species may provide 
clues to the formation of a molecule, such as more in the gas phase or more on grain surfaces, 
as in the case of dimethyl ether \citep{13C-DME_rot_det_2013}, among others.

The $^{12}$C/$^{13}$C ratio is particularly low in the vicinity of the Galactic center 
($\sim$20) \citep{isotopic_abundances,12C-13C_gradient_2005,13C-VyCN_2008}. Thus, it is not 
surprising that some $^{13}$C containing saturated organic molecules were detected soon after 
the main species and mainly toward the Galactic center source Sgr~B2. These are 
H$^{13}$C(O)NH$_2$ \citep{13C-FA_det_1978}, $^{13}$CH$_3$OH \citep{13C-MeOH_det_1979}, 
and CH$_3 ^{13}$CN \citep{MeCN-T_kin}. Both $^{13}$C containing CH$_3$CN isotopomers were 
detected in a line survey of Orion \citep{Orion-survey_13CH3CN}. Detections of molecules 
containing $^{13}$C were often hampered by the lack of laboratory data. In recent years, 
there have been several reports not only on the laboratory spectroscopy of $^{13}$C 
containing isotopologs, but also on their detections in space. These include ethyl 
cyanide \citep{13C-EtCN_rot_det_2007}, vinyl cyanide \citep{13C-VyCN_2008}, methyl 
formate \citep{13C-MeFo_rot_det_2009}, and dimethyl ether \citep{13C-DME_rot_det_2013}. 
The $^{13}$C containing ethanol isotopomers have also been studied \citep{13C-EtOH_rot_2012}, 
but only tentatively identified in a line survey of Sgr~B2(N) carried out with the IRAM 
30~m radio telescope \citep{SgrB2-survey_2013}. A firm detection has been achieved only 
very recently \citep{ROH_RSH_2015}. Further investigations of $^{13}$C containing organic 
molecules include acetaldehyde \citep{13C-CH3CHO_rot_2015}, whose detection has not been 
reported as far as we know, and methanol \citep{13C-MeOH_rot_2014}.

Methyl cyanide, CH$_3$CN, also known as acetonitrile or cyanomethane, was among the early 
molecules to be detected by radio astronomical means. \citet{det-MeCN} detected it almost 
40 years ago toward the massive star-forming regions Sgr A and B close to the 
Galactic center. The molecule was also detected in its $\varv _8 = 1$ excited vibrational 
state ($E_{\rm{vib}} = 525~K$ and even in $\varv _8 = 2$ and in $\varv _4 = 1$ 
($E_{\rm{vib}} = 1324~K$ \citep{SgrB2-survey_2013,deuterated_SgrB2N2_2015}.
Several rarer isotopic species have been detected. They include CH$_2$DCN 
in the hot core sources IRc2 in OMC1 and, tentatively, in G34.3 \citep{det-CH2DCN} 
and in Sgr~B2(N2) \citep{deuterated_SgrB2N2_2015}. Sgr~B2(N2) is to the north of 
Sgr~B2(N1), which is also known as the Large Molecule Heimat 
\citep{deuterated_SgrB2N2_2015}. Even $^{13}$CH$_3^{13}$CN was detected in Sgr~B2(N), 
first tentatively \citep{SgrB2-survey_2013} and, more recently, with confidence 
\citep{deuterated_SgrB2N2_2015}. In addition, 
\citet{SgrB2-survey_2013,deuterated_SgrB2N2_2015} identified transitions of 
CH$_3^{13}$CN and $^{13}$CH$_3$CN in their $\varv _8 = 1$ excited vibrational states.

Methyl cyanide was also among the first molecules to be studied by microwave spectroscopy 
\citep{MeCN_1st-MW}. In their rotational and rovibrational study of $\varv _8 = 0$, 1, 
and 2, \citet{MeCN_rot_2015} revealed and analyzed pronounced interactions between 
the last two states. In addition, they analyzed an interaction between the first two states. 
A few years earlier, some of us contributed extensive accounts of the ground vibrational 
states of several isotopic species of methyl cyanide, including CH$_3^{13}$CN, 
$^{13}$CH$_3$CN, and CH$_3$C$^{15}$N \citep{MeCN-isos_gs_rot_2009}. \citet{MeCN-14_15N} 
and \citet{MeCN-15N} reported the rotational spectrum of CH$_3$C$^{15}$N in its 
$\varv _8 = 1$ excited state up to $J = 8 - 7$ below 144~GHz. The only published rotational 
data of CH$_3^{13}$CN and $^{13}$CH$_3$CN in their $\varv _8 = 1$ states were provided 
by \citet{MeCN-_1-2x13_1988}, who measured transitions up to $J = 3 - 2$ below 56~GHz.

We have measured rotational spectra of methyl cyanide in natural isotopic composition 
both in wide and in more limited frequency windows up to 1.63~THz to provide 
new or updated catalog entries for various isotopologs of methyl cyanide, as well as 
for excited vibrational states. In the present article we analyzed these and new 
spectra to derive spectroscopic parameters of $^{13}$CH$_3$CN, CH$_3^{13}$CN, 
and CH$_3$C$^{15}$N in their $\varv _8 = 1$ excited vibrational states; as usual, 
unlabeled atoms refer to $^{12}$C and $^{14}$N. Preliminary results from this study
were used to identify lines of the $^{13}$C species in 3~mm molecular line surveys 
of Sgr~B2(N) with the IRAM 30~m telescope \citep{SgrB2-survey_2013} and with the 
Atacama Large Millimeter/submillimeter Array (ALMA) \citep{deuterated_SgrB2N2_2015}.

\section{Experimental details}
\label{exptl}

All measurements at the Universit{\"a}t zu K{\"o}ln were recorded at room temperature 
in static or very slow flow mode, employing a 7~m long double path Pyrex glass cell 
equipped with Teflon windows and having an inner diameter of 100~mm. The pressure 
of CH$_3$CN was in the range of 0.3$-$0.5~Pa. Source-frequency modulation was used 
with demodulation at $2f$, causing an isolated line to appear close to a second 
derivative of a Gaussian.

The $J = 3 - 2$ transitions of $^{13}$CH$_3$CN and CH$_3^{13}$CN in their $\varv _8 = 1$ 
excited vibrational states around 53.8 and 55.3~GHz, respectively, were recorded with 
an Agilent E8257D microwave synthesizer as source and a Schottky diode detector. 
In addition, $J = 6 - 5$ transitions of $^{13}$CH$_3$CN and CH$_3$C$^{15}$N around 107.4~GHz 
and the $J = 5 - 4$ and $6 - 5$ transitions of CH$_3^{13}$CN around 92.2 and 110.6~GHz, 
respectively, were recorded with essentially the same instrument. Source frequencies 
were generated using a Virginia Diodes, Inc. (VDI) tripler driven by the synthesizer 
mentioned above. The measurements were similar to the ones of 1,2-propanediol 
\citep{1-2-PD_rot_2014} and mono-deuterated ethanol \citep{D-EtOH_rot_2015} taken 
at lower frequencies.

We assigned uncertainties from 5 to 20~kHz to these lines because the lines were so 
narrow at these low frequencies and because of the good signal-to-noise ratios (S/Ns). 
Hyperfine structure caused by the $^{14}$N nucleus was resolved partially for several 
of these transitions.

The majority of the data were extracted from broad frequency scans taken with the 
JPL cascaded multiplier spectrometer \citep{JPL-spectrometer}. Generally, a multiplier 
chain source is passed through a one to two-meter path length flow cell and is detected 
by a silicon bolometer cooled to near 1.7~K. The cell is filled with a steady flow of 
reagent grade acetonitrile and the pressure and modulation are optimized to enable 
good S/Ns with narrow lineshapes. With a gas with very strong transitions, 
such as the $K < 7$ transitions of the main isotopolog of acetonitrile, the S/N 
was optimized for a higher $K$ transition (e.g., $K = 12$), such that the lower $K$ 
transitions exhibit saturated line profiles. This procedure enables a better dynamic 
range for the extraction of line positions for rare isotopologs and highly excited 
vibrational satellites. The frequency ranges covered most of the 400 to 1200~GHz region. 
Spectra around 1600~GHz were not considered in the present work because the ground 
state transitions of the three isotopic species were already too weak to be used 
in the final fits \citep{MeCN-isos_gs_rot_2009}. These measurements are similar 
to recent ones of mono-deuterated ethane \citep{EtD_rot_2015}, with one important 
difference being the very small dipole moment of the ethane isotopolog.

The efficiency of frequency multipliers usually changes strongly with frequency. 
In addition, recording conditions and sensitivities of detectors can have 
strong influences on the quality of the spectra. Uncertainties of 50 to 100~kHz 
were assigned to typical lines. Larger uncertainties of up to 200~kHz were assigned 
to weaker lines or lines that were not isolated, smaller uncertainties down to 
20~kHz for isolated lines with very good S/N and a very symmetric line shape.


\begin{figure}
  \includegraphics[angle=0,width=9cm]{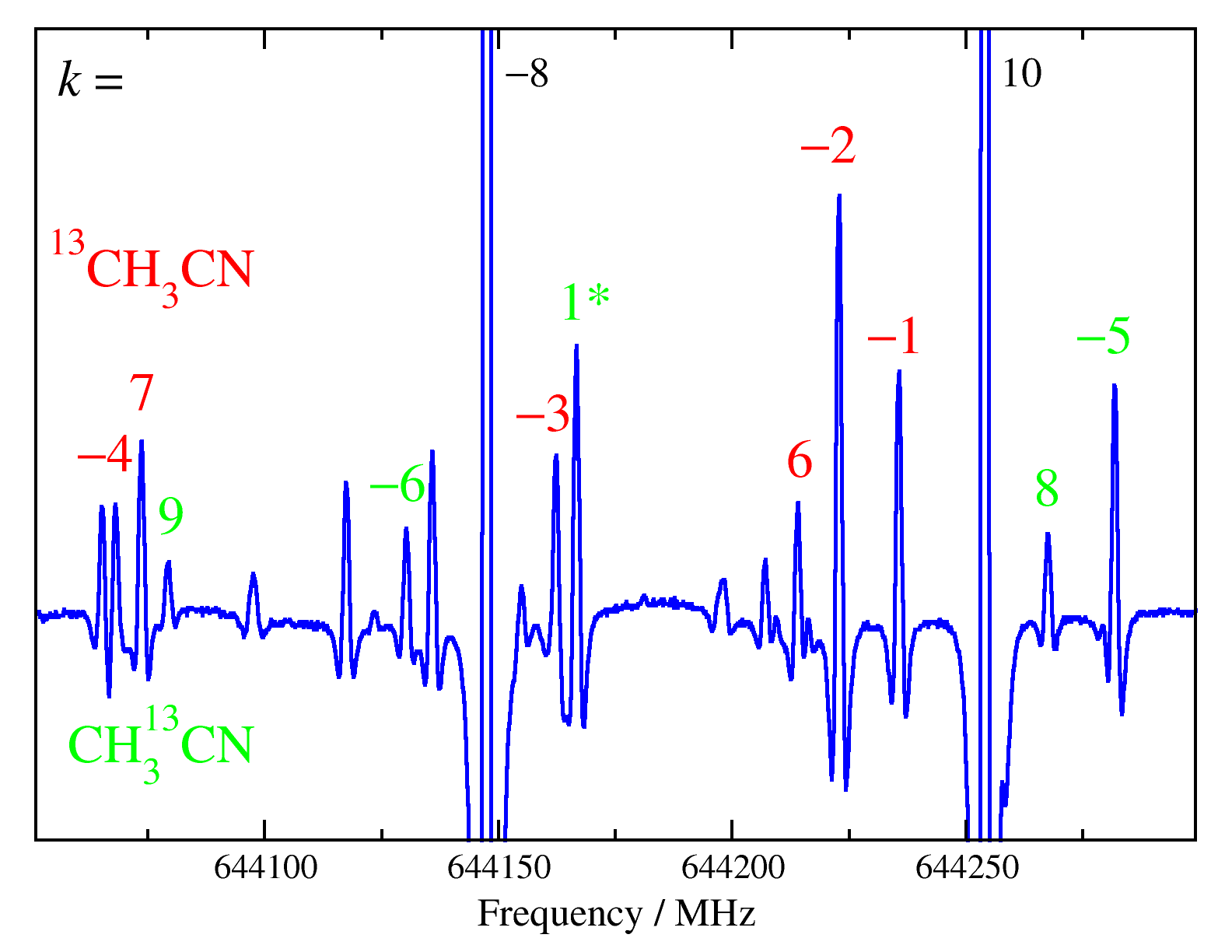}
  \caption{Section of the submillimeter spectrum of CH$_3$CN. Transitions of 
   $^{13}$CH$_3$CN ($J = 36 - 35$) and CH$_3^{13}$CN ($J = 35 - 34$) in their 
   $\varv _8 = 1$ excited vibrational states were labeled with their $k$ 
   quantum numbers. Spin-statistics cause $k = -2$, $-$5, and 7 to appear stronger 
   than expected, $1^{\ast}$ indicates the $A_2$ component of the asymmetry-like 
   split $k = +1$ transition, see section~\ref{r-a-d} for further details. 
   The clipped lines are the $J = 35 - 34$ $k = -8$ and $k = 10$ transitions 
   of the main isotopolog in $\varv _8 = 1$. The latter line is overlapping 
   the $A_1$ component of the $k = +1$ $J = 36 - 35$ line of CH$_3$C$^{15}$N, 
   $\varv _8 = 1$. The lines appear similar to second derivatives of a 
   Gaussian line-shape because of the 2$f$-modulation.}
\label{isos}
\end{figure}


\section{Results, analyses, and discussion}
\label{r-a-d}

The basic features of the rovibrational energy level structure of CH$_3$CN in its three 
lowest vibrational states ($\varv _8 \le 2$) have been given by \citet{MeCN_rot_2015}. 
The main isotopolog of methyl cyanide and the ones investigated in the present work 
are prolate symmetric rotors with $C_{\rm 3v}$ symmetry. Since the three light H 
atoms are the only ones not on the symmetry axis, the $A$ rotational parameter is 
much larger than $B$, $\sim$5.27~cm$^{-1}$ versus 0.307~cm$^{-1}$. The rotational 
energy increases therefore rapidly with $K$ and less so with $J$. The intensities 
of transitions involving higher $K$ levels drop quickly in intensity as a consequence.  
The rotational energies of the highest $K$ levels accessed in the present study 
correspond to $\sim$750~cm$^{-1}$ or $\sim$1080~K. The $\Delta K = 0$ selection 
rules for rotational transitions cause transitions with the same $J$ to occur in a 
comparatively narrow frequency range with the spacing increasing quite regularly 
with $K$ in the ground vibrational state. This is why rotational transitions 
of CH$_3$CN are frequently used to infer the temperature in the dense ISM. 
The spacing between two specific $K$ values also increases with $J$.

The lowest vibrationally excited state in CH$_3$CN is the doubly degenerate bending 
state $\varv _8 = 1$ with $E_{\rm{vib}} = 365$~cm$^{-1}$ or 525~K for the main isotopic 
species. Levels having the same $K$ ($> 0$) repel each other because of the Coriolis 
interaction, lifting the degeneracy. The $K$ levels in the $l = -1$ stack (i.e., 
$k = K \times \mathrm{sign}(l) < 0$) are shifted to higher energies, those in the 
$l = +1$ stack (i.e., $k > 0$) to lower energies. The Coriolis parameter $\zeta$ 
is $\sim$0.877, which is close to the limiting case of $\zeta = 1$, causing levels 
with $\Delta K = \Delta l = 2$ to be close in energy. The parameter $q_{22}$ causes 
these levels to repel each other more, because it induces interactions between 
levels differing in $\Delta K = \Delta l = 2$. The strength of the interaction 
depends strongly on $J$, as well as on $K$. The parameter $q_{22}$ is often 
abbreviated as $q$.

These two interactions together lead to a more irregular pattern of the 
transitions in $\varv _8 = 1$ with the same $J$ but differing $K$ compared 
with the quite regular pattern in $\varv = 0$. Those with $k \le 0$ occur 
for lower values of $J$ at decreasing frequencies with fairly regularly 
increasing spacing. Transitions with $k > 0$ occur less regularly to lower 
frequencies with $k = +2$ being higher in frequency than $K = 0$. Transitions 
with $\Delta K = \Delta l = 2$ are often close in frequency at higher $K$ 
in the absence of additional perturbations. The $q_{22}$ interaction leads 
to an asymmetry-like $A_1/A_2$ splitting for the $K = 1$, $l = +1$ level. 
The $A_1$ component is always highest in frequency for each $J$, whereas the 
$A_2$ component occurs near $k = -6$ for lower values of $J$. Figure~\ref{isos} 
shows a part of the rotational spectrum of CH$_3$CN, in particular with transition 
of the $^{13}$C isotopomers in their $\varv _8 = 1$ excited states.


\begin{table*}
\begin{center}
\caption{Spectroscopic parameters or differences ($\Delta$) thereof$^a$  of 
         methyl cyanide isotopologs within $\varv_8 = 1$.}
\label{parameter_v8_eq_1}
\begin{tabular}[t]{lr@{}lr@{}lr@{}lr@{}l}
\hline \hline 
Parameter & \multicolumn{2}{c}{CH$_3$CN$^b$} & \multicolumn{2}{c}{CH$_3^{13}$CN$^c$} &
 \multicolumn{2}{c}{$^{13}$CH$_3$CN$^c$}  & \multicolumn{2}{c}{CH$_3$C$^{15}$N$^c$}  \\
\hline
$E(8^1)$                    &      365&.024\,365\,(9)  &      357&.19             &      364&.56             &      362&.41             \\
$\Delta (A - B)$            &   $-$115&.930\,(26)      &   $-$115&.930            &   $-$115&.930            &   $-$115&.930            \\
$\Delta B$                  &       27&.530\,277\,(49) &       26&.404\,860\,(112)&       26&.688\,102\,(131)&       26&.865\,362\,(274)\\
$\Delta D_K \times 10^3$    &    $-$11&.46\,(48)       &    $-$11&.46             &    $-$11&.46             &    $-$11&.46             \\
$\Delta D_{JK} \times 10^3$ &        0&.987\,5\,(6)    &        0&.881\,1\,(24)   &        0&.945\,3\,(26)   &        1&.008\,2\,(50)   \\
$\Delta D_J \times 10^6$    &       95&.599\,(17)      &       92&.585\,(52)      &       90&.693\,(41)      &       90&.608\,(60)      \\
$\Delta H_K \times 10^6$    &       14&.9\,(22)        &       15&.               &       15&.               &       15&.               \\
$\Delta H_{KJ} \times 10^6$ &        0&.034\,1\,(22)   &        0&.034            &        0&.033            &        0&.033            \\
$\Delta H_{JK} \times 10^9$ &        2&.59\,(6)        &        2&.54             &        2&.44             &        2&.43             \\
$\Delta H_J \times 10^{12}$ &      315&.3\,(30)        &      307&.3\,(74)        &      278&.6\,(56)        &      293&.6\,(82)        \\
$\Delta L_J \times 10^{15}$ &     $-$2&.64\,(20)       &     $-$2&.54             &     $-$2&.35             &     $-$2&.30             \\
$\Delta eQq$                &     $-$0&.038\,7\,(19)   &     $-$0&.038\,7         &     $-$0&.038\,7         &         &n.a.            \\
$eQq\eta$                   &        0&.1519\,(113)    &        0&.1519           &        0&.1519           &         &n.a.            \\
$A\zeta$                    & 138\,656&.20\,(7)        & 139\,839&.99\,(39)       & 138\,857&.97\,(53)       & 138\,554&.93\,(105)      \\
$\eta_K$                    &       10&.332\,9\,(72)   &       10&.420            &       10&.347            &       10&.317            \\
$\eta_J$                    &        0&.390\,469\,(7)  &        0&.389\,261\,(24) &        0&.371\,764\,(23) &        0&.368\,215\,(34) \\
$\eta_{KK} \times 10^6$     &   $-$834&.\,(41)         &   $-$840&.               &   $-$835&.               &   $-$833&.               \\
$\eta_{JK} \times 10^6$     &    $-$34&.06\,(6)        &    $-$32&.42\,(44)       &    $-$32&.87\,(40)       &    $-$32&.27\,(60)       \\
$\eta_{JJ} \times 10^6$     &     $-$2&.359\,5\,(24)   &     $-$2&.275\,0\,(52)   &     $-$2&.147\,1\,(49)   &     $-$2&.173\,6\,(38)   \\
$\eta_{JKK} \times 10^9$    &        2&.59\,(17)       &        2&.6              &        2&.5              &        2&.5              \\
$\eta_{JJK} \times 10^9$    &        0&.509\,(6)       &        0&.542\,(69)      &        0&.425\,(70)      &        0&.478            \\
$q$                         &       17&.798\,44\,(2)   &       18&.212\,58\,(33)  &       16&.804\,42\,(35)  &       16&.860\,30\,(55)  \\
$q_{K} \times 10^3$         &     $-$2&.664\,5\,(111)  &     $-$2&.726            &     $-$2&.516            &     $-$2&.524            \\
$q_{J} \times 10^6$         &    $-$63&.842\,(14)      &    $-$64&.581\,(175)     &    $-$58&.745\,(164)     &    $-$58&.664\,(248)     \\
$q_{JK} \times 10^9$        &       93&.19\,(53)       &       95&.2              &       85&.4              &       85&.6              \\
$q_{JJ} \times 10^{12}$     &      311&.5\,(15)        &      304&.8\,(252)       &      301&.0\,(218)       &      257&.3\,(327)       \\
rms error                   &        0&.802            &        0&.808            &        0&.766            &        0&.826            \\
\hline
\end{tabular}
\end{center}
\tablefoot{
$^a$ All parameters in MHz units except $E(8^1)$ in cm$^{-1}$; the rms errors of the fits are unitless. 
Numbers in parentheses are one standard deviation in units of the least significant figures. 
Parameters without quoted uncertainties were estimated and kept fixed in the analyses; see also 
section~\ref{r-a-d}. 
$^b$ From \citet{MeCN_rot_2015}. 
$^c$ This work. Ground state parameters for corresponding isotopolog kept fixed to values in 
Table~\ref{ground-state-parameter}. Parameters in Table~\ref{interaction-parameter} were also used 
in the analyses.
}
\end{table*}


\begin{table*}
\begin{center}
\caption{Ground-state spectroscopic parameters$^a$ (MHz) of methyl cyanide isotopologs.}
\label{ground-state-parameter}
\renewcommand{\arraystretch}{1.10}
\begin{tabular}[t]{lr@{}lr@{}lr@{}lr@{}l}
\hline \hline 
Parameter & \multicolumn{2}{c}{CH$_3$CN$^b$} & \multicolumn{2}{c}{CH$_3^{13}$CN$^c$} &
 \multicolumn{2}{c}{$^{13}$CH$_3$CN$^c$}  & \multicolumn{2}{c}{CH$_3$C$^{15}$N$^c$}  \\
\hline
$(A - B)$                       & 148\,900&.103\,(66)      & 148\,904&.65              & 149\,165&.69              & 149\,176&.96               \\
$B$                             &   9\,198&.899\,167\,(11) &   9\,194&.349\,998\,(27)  &   8\,933&.309\,429\,(28)  &   8\,922&.038\,646\,(72)   \\
$D_K \times 10^3$               &   2\,830&.6\,(18)        &   2\,831&.                &   2\,831&.                &   2\,831&.                 \\
$D_{JK} \times 10^3$            &      177&.407\,87\,(25)  &      176&.673\,95\,(132)  &      168&.239\,71\,(130)  &      168&.938\,71\,(336)   \\
$D_J \times 10^6$               &   3\,807&.576\,(8)       &   3\,809&.737\,(37)       &   3\,624&.947\,(32)       &   3\,555&.251\,(45)        \\
$H_K \times 10^6$               &      164&.6\,(66)        &      165&.                &      165&.                &      165&.                 \\
$H_{KJ} \times 10^6$            &        6&.062\,0\,(14)   &        6&.004\,5\,(131)   &        5&.803\,0\,(141)   &        5&.647\,6\,(181)    \\
$H_{JK} \times 10^9$            &   1\,025&.69\,(15)       &   1\,017&.45\,(86)        &      927&.19\,(86)        &      951&.05\,(131)        \\
$H_J \times 10^{12}$            &   $-$237&.4\,(21)        &   $-$258&.4\,(61)         &   $-$273&.3\,(49)         &   $-$183&.1\,(67)          \\
$L_{KKJ} \times 10^{12}$        &   $-$444&.3\,(25)        &   $-$444&.                &   $-$431&.                &   $-$444&.                 \\
$L_{JK} \times 10^{12}$         &    $-$52&.75\,(51)       &    $-$49&.75\,(236)       &    $-$49&.66\,(266)       &    $-$49&.6                \\
$L_{JJK} \times 10^{12}$        &     $-$7&.901\,(32)      &     $-$7&.214\,(141)      &     $-$6&.887\,(128)      &     $-$7&.057\,(182)       \\
$L_J \times 10^{15}$            &     $-$3&.10\,(17)       &     $-$3&.10              &     $-$2&.76              &     $-$2&.74               \\
$P_{JK} \times 10^{15}$         &        0&.552\,(68)      &        0&.55              &        0&.51              &        0&.50               \\
$P_{JJK} \times 10^{18}$        &       55&.3\,(22)        &       55&.                &       49&.                &       49&.                 \\
$eQq$                           &     $-$4&.223\,08\,(107) &     $-$4&.218\,28\,(176)  &     $-$4&.218\,30\,(197)  &         &n.a.              \\
$C_{bb} \times 10^3$            &        1&.845\,(90)      &        1&.844             &        1&.792             &         &                  \\
$(C_{aa} - C_{bb}) \times 10^3$ &     $-$1&.15\,(31)       &     $-$1&.15              &     $-$1&.10              &         &                  \\
\hline
\end{tabular}
\end{center}
\tablefoot{
$^a$ Numbers in parentheses are one standard deviation in units of the least significant figures. 
Parameters without quoted uncertainties have been estimated from the main isotopic species and were kept fixed in the fits; 
see section~\ref{r-a-d}.
$^b$ From \citet{MeCN_rot_2015}. 
$^c$ From \citet{MeCN-isos_gs_rot_2009}; adjusted in this work to account for slight changes in the parameters of the main isotopolog 
from that work compared to those in \citet{MeCN_rot_2015}. 
}
\end{table*}


\begin{table*}
\begin{center}
\caption{Low-order spectroscopic parameters or differences ($\Delta$) thereof$^a$  of 
         methyl cyanide describing the Fermi interaction between $\varv_8 = 1$ and 
         $\varv_8 = 2$.}
\label{interaction-parameter}
\begin{tabular}[t]{lr@{}lr@{}lr@{}lr@{}l}
\hline \hline 
Parameter & \multicolumn{2}{c}{CH$_3$CN$^b$} & \multicolumn{2}{c}{CH$_3^{13}$CN$^c$} &
 \multicolumn{2}{c}{$^{13}$CH$_3$CN$^c$}  & \multicolumn{2}{c}{CH$_3$C$^{15}$N$^c$}  \\
\hline
$E(8^{2^0})$                              &      716&.750\,42\,(13)   &      701&.10      &      715&.82      &      711&.53      \\
$\Delta (A - B)$                          &   $-$187&.404\,(18)       &   $-$187&.4       &   $-$187&.4       &   $-$187&.4       \\
$\Delta B$                                &       54&.057\,316\,(111) &       51&.856     &       52&.411\,5  &       52&.759\,1  \\
$E(8^{2^2})$                              &      739&.148\,225\,(56)  &      723&.01      &      738&.19      &      733&.76      \\
$\Delta (A - B)$                          &   $-$259&.956\,(122)      &   $-$260&.0       &   $-$260&.0       &   $-$260&.0       \\
$\Delta B$                                &       54&.502\,729\,(70)  &       52&.283\,3  &       52&.843\,35 &       53&.193\,85 \\
$A\zeta$                                  & 138\,656&.042\,(102)      & 139\,829&.5       & 138\,858&.        & 138\,555&.        \\
$q$                                       &       17&.729\,857\,(138) &       18&.141\,2  &       16&.738\,2  &       16&.794\,9  \\
$F(8^{\pm1},8^{2,\mp2})$                  &  53\,157&.7\,(33)         &  54\,567&.\,(744) &  51\,647&.\,(1303)&  51\,298&.\,(2967)\\
$F_K(8^{\pm1},8^{2,\mp2})$                &      $-$6&.               &      $-$6&.       &      $-$6&.       &      $-$6&.       \\
$F_J(8^{\pm1},8^{2,\mp2})$                &      $-$0&.369\,89\,(44)  &      $-$0&.370    &      $-$0&.359    &      $-$0&.359    \\
$F_{JJ}(8^{\pm1},8^{2,\mp2}) \times 10^6$ &         1&.681\,(87)      &         1&.70     &         1&.58     &         1&.58     \\
\hline
\end{tabular}
\end{center}
\tablefoot{
$^a$ All parameters in MHz units except $E(8^{2^0})$ and $E(8^{2^2})$ in cm$^{-1}$. 
Numbers in parentheses are one standard deviation in units of the least significant figures. 
Parameters without quoted uncertainties were estimated and kept fixed in the analyses; see also 
section~\ref{r-a-d}. 
$^b$ From \citet{MeCN_rot_2015}. 
$^c$ This work. Ground state parameters for corresponding isotopolog kept fixed to values in 
Table~\ref{ground-state-parameter}. 
}
\end{table*}


Spin statistics involving the three equivalent H atoms lead to $A$ and $E$ symmetry levels 
with $ortho$ and $para$ spin states, respectively. The $K$ levels with $K - l \equiv 0$~mod~3 
have $A$ symmetry with twice the spin weight of the $E$ symmetry $K$ levels with the important 
exception of $K = 0$ in vibrational (sub-) states with $l \equiv 0$~mod~3, in particular 
in the ground vibrational state. If the $K$ levels show $A_1/A_2$ splitting (e.g., $k = +1$ 
in $\varv _8 = 1$ of the present CH$_3$CN isotopologs), each level carries half the spin weight, 
which means they appear in intensity as if they were $E$-level transitions.

The large dipole moment of methyl cyanide (3.92197~(13)~D, \citealt{MeCN-dipole}) leads 
to a very strong rotational spectrum, such that CH$_2$DCN and even $^{13}$CH$_3^{13}$CN 
are observable in natural isotopic composition under favorable conditions 
\citep{MeCN-isos_gs_rot_2009}. The dipole moment was also determined for CH$_3$C$^{15}$N, 
yielding an essentially identical value (3.9256~(7)~D,  \citealt{MeCN-15-nu4_Stark_1984}). 
Dipole moments have also been determined for this isotopic species in excited vibrational 
states \citep{MeCN-15-nu4_Stark_1984,MeCN-15-nu4HB_Stark_1984}. Excitation to $\varv _8 = 1$ 
leads to a value of 3.9073~(13)~D.

The terrestrial $^{12}$C/$^{13}$C and $^{14}$N/$^{15}$N ratios are about 90 and 270, 
respectively \citep{composition_elements_2009}. Transitions in the $\varv _8 = 1$ 
excited state are weaker by almost a factor of six than those of the ground vibrational 
state at room temperature. Thus, CH$_3$C$^{15}$N transitions in $\varv _8 = 1$ are 
more than five times stronger than $^{13}$CH$_3^{13}$CN transitions in $\varv = 0$.
Initially, we searched for transitions of CH$_3$C$^{15}$N because attempts to assign 
$^{13}$C isotopomers in their $\varv _8 = 1$ states in astronomical data 
\citep{SgrB2-survey_2013,deuterated_SgrB2N2_2015} based on the previous laboratory 
data \citep{MeCN-_1-2x13_1988} already displayed deviations of up to 2~MHz in the 
3~mm wavelength region.

We adjusted the previous $\varv _8 = 1$ spectroscopic parameters of CH$_3$C$^{15}$N 
\citep{MeCN-14_15N,MeCN-15N} to those used for the main isotopic species 
\citep{MeCN_rot_2015} and estimated higher order parameters for the $^{15}$N species 
as described below. After fitting the previously reported transition frequencies 
with their reported uncertainties and creating predictions for higher frequencies, 
assignments could be made rather easily in the 446$-$483~GHz region ($J'' = 24 - 26$). 
Subsequently, we estimated spectroscopic parameters for the CH$_3$CN isotopomers 
containing one $^{13}$C isotope in their $\varv _8 = 1$ states from their ground-state 
parameters \citep{MeCN-isos_gs_rot_2009}, along with the CH$_3$CN and CH$_3$C$^{15}$N 
parameters in $\varv _8 = 0$ and 1. These assignments could be made relatively easily 
in essentially the same frequency window.

Subsequently, we increased the assignments in several steps to higher $J$ and $K$ 
values for all three isotopologs. Eventually, the $J$ range covered 24 to 66 
(23 to 64 for CH$_3^{13}$CN for which the $B$ value is almost the same as for 
the main species). Around 20 transition frequencies could be assigned for each $k$ 
and each of the two $^{13}$C isotopomers between $k = -6$ and +10 with assigments 
decreasing to higher $K$ because of decreasing intensity. The number of assigned 
transition frequencies for each $k$ of CH$_3$C$^{15}$N were very similar for the 
lowest energy $k$ values, but decreased earlier because of the lower abundance 
of the isotopolog compared to the two with one $^{13}$C isotope. Assignments reached 
$k = +13$ and initially $k = -9$ for all three isotopic species. Based on intensities, 
we expected to be able to assign at least some transitions with $k = -10$ and $-11$, 
but these were not found to be sufficiently close to the predictions. We suspected that 
these rather weak transitions were perturbed by $K = 10$ and 11 of $\varv _8 = 2$, 
$l = +2$ via a strong $\Delta \varv_8 = \pm 1$, $\Delta K = 0$, $\Delta l = \pm 3$ 
Fermi resonance with the strongest effects at $K = 14$ for the main isotopic species 
\citep{MeCN_rot_2015}. Including estimates of low-order spectroscopic parameters 
for $\varv _8 = 2$ and the $\varv _8 = 1$/$\varv _8 = 2$ interaction parameters 
into the fit, we were able to assign some transitions with $k = -10$ and $-11$ 
for all three isotopologs that displayed perturbations of up to 1~MHz and around 
2~MHz, respectively. In the late stages of the project, we also made assignments 
at millimeter wavelengths for the three isotopic species. These covered $J'' = 2$, 4 
and 5 transitions for CH$_3^{13}$CN, $J'' = 2$ and 5 for $^{13}$CH$_3$CN, and 
$J'' = 5$ for CH$_3$C$^{15}$N. Hyperfine structure caused by the $^{14}$N nucleus 
was resolved partially for several of the transitions in the millimeter-wave range.

Prediction and fitting of the rotational spectra were carried out with the Spcat/Spfit 
program suite \citep{spfit_1991}. In the CH$_3$C$^{15}$N fits, we used the previously 
recorded laboratory data \citep{MeCN-14_15N,MeCN-15N}, except for the $J'' = 5$ 
transition frequencies, which were replaced by our more accurate data. 
The $\varv _8 = 1$ and $\varv _8 = 2^0$ vibrational energies were taken from 
\citet{MeCN-FF_1978} and corrected for the differences with respect to values determined 
for the main species by \citet{MeCN_rot_2015}. The correction was only 0.02~cm$^{-1}$ 
in the latter case because the sharp $Q$ branch of the parallel $2\nu _8^0$ band 
permit accurate determinations of the vibrational energies. The absence of a single 
sharp $Q$ branch in the perpendicular $\nu _8$ band led to a larger correction of 
0.31~cm$^{-1}$. Scaling of the vibrational energies of $\varv _8 = 2^0$ with 
the ratio $E(\varv _8 = 2^2)/E(\varv _8 = 2^0)$ of the main isotopolog was used 
to estimate $E(\varv _8 = 2^2)$ values for the three minor isotopic species.

The purely axial parameters $A$ (or $A-B$), $D_K$, $H_K$, etc., of a symmetric top 
molecule cannot be determined by rotational spectroscopy in the absence of perturbations. 
Moreover, rovibrational spectroscopy is only able to determinethe differences 
with respect to the ground vibrational state unless ground-state $\Delta K = 3$ 
loops are formed from rovibrational spectra involving two degenerate states, 
as was done initially for CH$_3$CN \citep{MeCN-A}, or through perturbations, or 
through a combination of both, which we used in our study on $\varv _8 \le 2$ states 
of the main isotopolog \citep{MeCN_rot_2015}.

The ground-state spectroscopic parameters of the minor isotopic species were initially 
taken from \citet{MeCN-isos_gs_rot_2009}. Because of slight changes in the ground-state 
parameters of CH$_3$CN in our recent study \citep{MeCN_rot_2015}, we adjusted the 
purely axial parameters, as well as the estimates of some higher order parameters, 
and redetermined the remaining ground-state spectroscopic parameters from the data 
in \citet{MeCN-isos_gs_rot_2009}. Initial $\varv _8 = 1$ spectroscopic parameters of 
CH$_3^{13}$CN and $^{13}$CH$_3$CN or changes thereof, as well as higher order values 
of CH$_3$C$^{15}$N, were derived from corresponding values of CH$_3$CN by scaling them 
with appropriate powers of the ratio of the $B$ rotational constants, as was done 
earlier for the ground vibrational data \citep{MeCN-isos_gs_rot_2009}. Later, the 
distortion corrections $\eta$ to $A\zeta$ were also scaled with the $A\zeta$ ratio, 
and the distortion corrections to $q$ were scaled equivalently. Later, lower order 
spectroscopic parameters of $\varv _8 = 2$ were estimated in a similar way. 
The Fermi parameter $F$, which described the $\Delta \varv_8 = \pm 1$, $\Delta K = 0$, 
$\Delta l = \pm 3$ interaction between $\varv _8 = 1$ and 2, was assumed to be identical 
for all isotopologs. The hyperfine structure was reproduced well by the ground-state 
parameters of the respective isotopologs combined with the changes from the ground 
to the $\varv _8 = 1$ state of the main isotopic species.

Throughout, we tried to fit the rotational spectra of the three methyl cyanide isotopologs 
with as few spectroscopic parameters being floated as possible. Starting from the lowest 
order parameters, we searched at each intermediate fit for the parameter that reduced 
the rms error (also known as the reduced or weighted $\chi$ squared) of the fit most. 
The rms error of the fit is a measure of the quality of the fit, and it should be close 
to 1.0 ideally, preferably slightly smaller. We tried to avoid floating parameters 
that changed from the initial value by too much.

Initially, our fits did not include any distortion corrections to $F$. However, we were 
unable to reproduce transitions with $k = -10$ and $-11$ satisfactorily. Floating $F$ yielded 
relatively small changes from the initial values compared to floating other parameters, 
but the values were increased by about 15\,\% to 20\,\%. Trial fits, with fixed $F_J$ values 
scaled with the ratio of the $B$ rotational constants included, reduced the values of $F$ 
to around that of the main isotopic species. Even though including $F_K$ and $F_{JJ}$ 
had effects within the uncertainties of $F$, they were retained in the fits.
The final set of $\varv _8 = 1$ specroscopic parameters of the three methyl cyanide 
isotopic species is given in Table~\ref{parameter_v8_eq_1}, along with values for 
the main species from \citet{MeCN_rot_2015}. The adjusted ground-state parameters are 
provided in Table~\ref{ground-state-parameter}. These parameters were kept fixed in 
all $\varv _8 = 1$ fits. Finally, the low-order $\varv _8 = 2$ parameters are summarized 
in Table~\ref{interaction-parameter}, along with those describing the Fermi interaction 
between $\varv _8 = 1$ and 2.

The rms errors of all fits are around 0.8, so the experimental transition frequencies 
have been reproduced within uncertainties on average, and the uncertainties even appear 
to be slightly conservative. The $\varv _8 = 1$ parameters in Table~\ref{parameter_v8_eq_1} 
are quite similar among the isotopic species, in particular those of $^{13}$CH$_3$CN 
and CH$_3$C$^{15}$N, which have rather similar vibrational energies and values of 
$\Delta B$, $A\zeta$, and $q$. The ground-state parameters of three minor isotopic 
species change only slightly with respect to our previous values because of relatively 
modest changes in the estimates of $P_{JK}$ and somewhat larger changes in $L_J$. 
The change in $H_K$ has no effect on the not purely axial parameters. The only floated 
parameter in Table~\ref{interaction-parameter} is the main Fermi term $F$. The values 
of the minor isotopic species differ slightly from those of the main species. 
Assumptions on the purely axial parameters, as well as the truncation of the 
$\varv _8 = 2$ parameters, may have non-negligible effects on its values. 
On the other hand, \citet{MeCN_rot_2015} pointed out that $F$ appeared to scale 
roughly with $q$ for the isoelectronic molecules CH$_3$CN, CH$_3$CCH 
\citep{CH3CCH_rot_vib_2004} and CN$_3$NC \citep{CH3NC_rot_vib_2011}. 
Interestingly, $q$ and $F$ of CH$_3^{13}$CN are both slightly greater than the values 
of the main species, whereas both are less by very similar amounts for $^{13}$CH$_3$CN 
and CH$_3$C$^{15}$N, and, remarkably, basically identical with $q = 16.78926~(19)$~MHz 
and $F = 51745~(3)$~MHz of CH$_3$CCH \citep{CH3CCH_rot_vib_2004}. However, the agreement 
with the CH$_3$CCH values may be coincidental.

\section{Conclusion}
\label{conclusion}

We have analyzed rotational transitions for three minor isotopologs of methyl cyanide in 
their $\varv _8 = 1$ excited vibrational states up to 1.2~THz and combined these data with 
existing ones in the case of CH$_3$C$^{15}$N. While the prospects of detecting such lines 
for this isotopic species are uncertain at present, transitions pertaining to $^{13}$CH$_3$CN 
and CH$_3^{13}$CN have already been identified in astronomical spectra with the help of 
preliminary results from this study \citep{SgrB2-survey_2013,deuterated_SgrB2N2_2015}.

Predictions generated from the present data should be sufficient for observations 
with ALMA, and even more so with other arrays or single-dish radio telescopes. 
These predictions will be available in the catalog 
section\footnote{website: https://cdms.astro.uni-koeln.de/classic/entries/, 
see also https://cdms.astro.uni-koeln.de/classic/}  
of the Cologne Database for Molecular 
Spectroscopy\footnote{website: https://cdms.astro.uni-koeln.de/}
\citep{CDMS_1,CDMS_2}. The complete line, parameter, and fit files 
will be deposited in the Spectroscopy Data section of the CDMS. 
Updated or new JPL catalog \citep{JPL-catalog} entries\footnote{website:  
http://spec.jpl.nasa.gov/ftp/pub/catalog/catdir.html, see also 
http://spec.jpl.nasa.gov/} will also be available.

The very good S/N of our spectra should permit analyses of rotational 
transitions of $^{13}$CH$_3$CN and CH$_3^{13}$CN, possibly even of 
CH$_3$C$^{15}$N in their $\varv _8 = 2$ and $\varv _4 = 1$ excited 
vibrational states. Such transitions may well be detectable in astronomical 
spectra in case of the $^{13}$C containing isotopomers.

The use of enriched samples will permit accessing energy levels having even higher 
$J$ and $K$ quantum numbers, which will yield an improved description of the 
$\varv _8 = 1$ and 2 interactions. However, such a study will also require recording 
and analyses of $\nu _8$ and $2\nu _8$ infrared spectra.


\begin{acknowledgements}
The measurements in K{\"o}ln were supported by the Deutsche Forschungsgemeinschaft 
(DFG) through the collaborative research grant SFB~956, project area B3. 
We would like to thank Dr. Bernd Vowinkel for making Schottky detectors available 
for our measurements. 
The portion of this work, which was carried out at the Jet Propulsion Laboratory, 
California Institute of Technology, was performed under contract with the 
National Aeronautics and Space Administration (NASA). 
\end{acknowledgements}



\begin{thebibliography}{}


\bibitem[Anttila et al.(1993)]{MeCN-A} 
Anttila, R., Horneman, V.-M., Koivusaari, M. \& Paso, R. 
1993, J. Mol. Spectrosc., 157, 198

\bibitem[Bauer \& Maes(1969)]{MeCN-14_15N} 
Bauer, A., \& Maes, S. 
1969, J. Phys. (Paris), 30, 169 

\bibitem[Bauer et al.(1975)]{MeCN-15N} 
Bauer, A., Tarrago, G., \& Remy, A. 
1975, J. Mol. Spectrosc., 58, 111 

\bibitem[Belloche et al.(2009)]{det-PrCN_EtFo} 
Belloche, A., Garrod, R.~T., M{\"u}ller, H.~S.~P., et al. 
2009, \aap, 499, 215

\bibitem[Belloche et al.(2013)]{SgrB2-survey_2013} 
Belloche, A., M{\"u}ller, H.~S.~P., Menten, K.~M., Schilke, P., \& Comito, C. 
2013, \aap, 559, A47 

\bibitem[Belloche et al.(2014)]{i-PrCN_det_2014} 
Belloche, A., Garrod, R.~T., M{\"u}ller, H.~S.~P., \& Menten, K.~M. 
2014, Science, 345, 1584 

\bibitem[Belloche et al.(2015)]{deuterated_SgrB2N2_2015} 
Belloche, A., M{\"u}ller, H.~S.~P., Garrod, R.~T., \& Menten, K.~M. 
2016, \aap, 587, A91

\bibitem[Berglund \& Wieser(2011)]{composition_elements_2009}
Berglund, M; \& Wieser, M.~E.\
2011, Pure Appl. Chem., 83, 397

\bibitem[Bossa et al.(2014)]{1-2-PD_rot_2014} 
Bossa, J.-B., Ordu, M.~H., M{\"u}ller, H.~S.~P., Lewen, F., \& Schlemmer, S. 
2014, \aap, 570, A12 

\bibitem[Bouchez et al.(2012)]{13C-EtOH_rot_2012} 
Bouchez, A., Walters, A., M{\"u}ller, H.~S.~P., et al. 
2012, J. Quant.  Spectrosc. Radiat. Transfer, 113, 1148 

\bibitem[Carvajal et al.(2009)]{13C-MeFo_rot_det_2009} 
Carvajal, M., Margul{\`e}s, L., Tercero, B., et al. 
2009, \aap, 500, 1109 

\bibitem[Cummins et al.(1983)]{MeCN-T_kin} 
Cummins, S.~E., Green, S., Thaddeus, P., \& Linke, R.~A.\ 
1983, \apj, 266, 331 

\bibitem[Daly et al.(2015)]{EtD_rot_2015} 
Daly, A.~M., Drouin, B.~J., Groner, P., Yu, S., \& Pearson, J.~C. 
2015, J. Mol. Spectrosc., 307, 27 

\bibitem[Demyk et al.(2007)]{13C-EtCN_rot_det_2007} 
Demyk, K., M{\"a}der, H., Tercero, B., et al. 
2007, \aap, 466, 255 

\bibitem[Drouin et al.(2005)]{JPL-spectrometer} 
Drouin, B.~J., Maiwald, F.~W., \& Pearson, J.~C. 
2005, Rev. Sci. Instr., 76, 093113

\bibitem[Duncan et al.(1978)]{MeCN-FF_1978} 
Duncan, J.~L., McKean, D.~C., Tullini, F., Nivellini, G.~D., \& Perez Pe{\~n}a, J. 
1978, J. Mol. Spectrosc., 69, 123 

\bibitem[Gadhi et al.(1995)]{MeCN-dipole}
Gadhi, J., Lahrouni, A., Legrand, J., \& Demaison, J. 
1995, J. Chim. Phys. Phys-Chim. Biol., 92, 1984

\bibitem[Gerin et al.(1992)]{det-CH2DCN} 
Gerin, M., Combes, F., Wlodarczak, et al.
1992, \aap, 259, L35 

\bibitem[Gottlieb et al.(1979)]{13C-MeOH_det_1979} 
Gottlieb, C.~A., Ball, J.~A., Gottlieb, E.~W., \& Dickinson, D.~F. 
1979, \apj, 227, 422 

\bibitem[Koerber et al.(2013)]{13C-DME_rot_det_2013} 
Koerber, M., Bisschop, S.~E., Endres, C.~P., et al. 
2013, \aap, 558, A112 

\bibitem[J{\o}rgensen et al.(2012)]{IRAS16293_glycald_2012} 
J{\o}rgensen, J.~K., Favre, C., Bisschop, S.~E., et al. 
2012, \apjl, 757, L4 

\bibitem[Lazareff et al.(1978)]{13C-FA_det_1978} 
Lazareff, B., Lucas, R., \& Encrenaz, P. 
1978, \aap, 70, L77 

\bibitem[Margul{\`e}s et al.(2015)]{13C-CH3CHO_rot_2015} 
Margul{\`e}s, L., Motiyenko, R.~A., Ilyushin, V.~V., \& Guillemin, J.~C. 
2015, \aap, 579, A46 

\bibitem[Maury et al.(2014)]{NGC1333-IRAS2A_survey_2014} 
Maury, A.~J., Belloche, A., Andr{\'e}, P., et al. 
2014, \aap, 563, L2 

\bibitem[Milam et al.(2005)]{12C-13C_gradient_2005} 
Milam, S.~N., Savage, C., Brewster, M.~A., Ziurys, L.~M., \& Wyckoff, S. 
2005, \apj, 634, 1126 

\bibitem[Mito et al.(1984a)]{MeCN-15-nu4_Stark_1984}  
Mito, A., Sakai, J., \& Katayama, M. 
1984a, J. Mol. Spectrosc., 103, 26

\bibitem[Mito et al.(1984b)]{MeCN-15-nu4HB_Stark_1984}  
Mito, A., Sakai, J., \& Katayama, M. 
1984b, J. Mol. Spectrosc., 105, 410

\bibitem[M{\"u}ller et al.(2001)]{CDMS_1}
M{\"u}ller, H.~S.~P., Thorwirth, S., Roth, D.~A.,
\& Winnewisser, G.
2001, A\&A, 370, L49

\bibitem[M{\"u}ller et al.(2005)]{CDMS_2}
M{\"u}ller, H.~S.~P., Schl{\"o}der, F., Stutzki, J.,
\& Winnewisser, G.
2005, J. Mol. Struct, 742, 215

\bibitem[M{\"u}ller et al.(2008)]{13C-VyCN_2008} 
M{\"u}ller, H.~S.~P., Belloche, A., Menten, K.~M., et. al.
2008, J. Mol. Spectrosc., 251, 319 

\bibitem[M{\"u}ller et al.(2009)]{MeCN-isos_gs_rot_2009} 
M{\"u}ller, H.~S.~P., Drouin, B.~J., \& Pearson, J.~C. 
2009, \aap, 506, 1487 

\bibitem[M{\"u}ller et al.(2015)]{MeCN_rot_2015} 
M{\"u}ller, H.~S.~P., Brown, L.~R., Drouin, B.~J., et al. 
2015a, J. Mol. Spectrosc., 312, 22 

\bibitem[M{\"u}ller et al.(2016)]{ROH_RSH_2015}
M{\"u}ller, H.~S.~P., Belloche, A., Xu, L.-H., et al. 
2016, \aap, 587, A92

\bibitem[Pickett(1991)]{spfit_1991} 
Pickett, H.~M. 
1991, J. Mol. Spectrosc., 148, 371

\bibitem[Pickett et al.(1998)]{JPL-catalog} 
Pickett, H.~M., Poynter, R.~L., Cohen, E.~A., et al. 
1998, J. Quant.  Spectrosc. Radiat. Transfer, 60, 883

\bibitem[Pineda et al.(2012)]{IRAS16293-ALMA-SV_2012} 
Pineda, J.~E., Maury, A.~J., Fuller, G.~A., et al. 
2012, \aap, 544, L7 

\bibitem[Pracna et al.(2004)]{CH3CCH_rot_vib_2004} 
Pracna, P., M{\"u}ller, H.~S.~P., Klee, S., \& Horneman, V.-M. 
2004, Mol. Phys., 102, 1555 

\bibitem[Pracna et al.(2011)]{CH3NC_rot_vib_2011} 
Pracna, P., Urban, J., Votava, O., et al. 
2011, Mol. Phys., 109, 2237 

\bibitem[Ring et al.(1947)]{MeCN_1st-MW} 
Ring, H., Edwards, H., Kessler, M., \& Gordy, W. 
1947, Phys. Rev., 72, 1262

\bibitem[Solomon et al.(1971)]{det-MeCN} 
Solomon, P.~M., Jefferts, K.~B., Penzias, A.~A., \& Wilson, R.~W.\ 
1971, \apjl, 168, L107 

\bibitem[Sutton et al.(1985)]{Orion-survey_13CH3CN} 
Sutton, E.~C., Blake, G.~A., Masson, C.~R., \& Phillips, T.~G. 
1985, \apjs, 58, 341 

\bibitem[Tam et al.(1988)]{MeCN-_1-2x13_1988} 
Tam, H., An, I., \& Roberts, J. A.
1988, J. Mol. Spectrosc., 129, 202

\bibitem[Tercero et al.(2015)]{search_EME_n-PrOH_2015} 
Tercero, B., Cernicharo, J., L{\'o}pez, A., et al. 
2015, \aap, 582, L1

\bibitem[Walters et al.(2015)]{D-EtOH_rot_2015} 
Walters, A., Sch{\"a}fer, M., Ordu, M.~H., et al. 
2015, J. Mol. Spectrosc., 314, 6 

\bibitem[Wilson \& Rood(1994)]{isotopic_abundances} 
Wilson, T.~L., \& Rood, R. 
1994, \araa, 32, 191

\bibitem[Wyrowski et al.(2003)]{CRL618-model_2003} 
Wyrowski, F., Schilke, P., Thorwirth, S., Menten, K.~M., \& Winnewisser, G. 
2003, \apj, 586, 344 

\bibitem[Xu et al.(2014)]{13C-MeOH_rot_2014} 
Xu, L.-H., Lees, R.~M., Hao, Y., et al. 
2014, J. Mol. Spectrosc., 303, 1 


\end{thebibliography}
\end{document}